\documentclass[12pt,preprint]{aastex}
\usepackage{amssymb}
\usepackage{color}

\shorttitle{Planetesimal Accretion in Binary Systems} \shortauthors{Xie & Zhou}
\begin{document}

\title{PLANETESIMAL ACCRETION IN BINARY SYSTEMS: ROLE OF THE  COMPANION'S ORBITAL INCLINATION}
\author{Ji-Wei Xie and Ji-Lin Zhou}
\affil{Department of Astronomy, Nanjing University, 210093
Nanjing, China} \email{J.XIE@ASTRO.UFL.EDU or XIEJIWEI@Gmail.com}

\begin{abstract}
Recent observations show that planet can reside in close binary
systems with stellar separation of only $\sim$20 AU. However,
planet formation in such close binary systems is a challenge to
current theory. One of the major theoretical problems occurs in the intermediate
stage$-$planetesimals accretion into planetary embryos$-$during
which the companion's perturbations can stir up the relative
velocites($\triangle V$) of planetesimals and thus slow down or
even cease their growth. Recent studies shown conditions could be even worse for accretion if the gas disk evolution was included.
However, all previous studies assumed a 2-dimentional (2D) disk and a coplanar binary orbit. Extending previous studies by including a 3D gas disk and an inclined binary orbit with small relative inclination of $i_B$=0.1-5 deg, we numerically investigate the conditions for planetesimal accretion at 1-2 AU, an extension of the habitable zone($\sim$1-1.3 AU), around $\alpha$ Centauri A in this paper. Inclusion of the binary inclination leads to: (1) differential orbital phasing is realized in the 3D space, and thus different-sized bodies are separated from each other; (2) total impact rate is lower, and impacts mainly occur between similar-sized bodies; (3) accretion is more favored, but the balance between accretion and erosion remains uncertain,
and the ``possible accretion region" extends up to 2AU when assuming an optimistic $Q^*$(critical specific energy that leads to catastrophic fragmentation); and (4) impact velocities ($\triangle V$) are significantly reduced but still much larger than their escape velocities, which infers that planetesimals grow by means of type II runaway mode. As a conclusion, inclusion of a small binary inclination is a promising mechanism that favors accretion, opening a possibility that planet formation in close binary systems can go through the difficult stage of planetesimals accretion into planetary embryos. 
\end{abstract}

\clearpage

\section{INTRODUCTION}
As of today, more than 300 extrasolar planets have been found
around main sequence stars,  and about 33 of them are known to
reside in binary systems(Eggenberger et al. 2004; Konachi 2005;
Mugrauer et al. 2005). As solar-type stars usually born in binary
or multiple systems(Duquennoy \& Mayor 1991; Mathieu et al. 2000;
Duch{\^e}ne et al. 2007), more and more planets are expected in
binary systems with the improvement of the detecting techniques.
In most detected planet-bearing binary systems,
planet resides around one star with the other companion acting as a
perturber on an outer orbit(so called S-Type orbit). The orbital separations of the binary are found in a wide range from thousands to only a few tens of AU. Three systems$-$ GJ86 (Queloz et al. 2000), $\gamma$ Cephei (Hatzes et al. 2003), and HD 41004 (Zucker et al. 2004)$-$ have
stellar semimajor axes of only $\sim$20 AU. Moreover, the star HD
188753A, which might host a Jupiter-mass planet (supported by Konachi 2005, but challengaed by Eggenberger et al. 2007) on a 3.35 day S-Type
orbit, has a companion with a close periastron distance of only
$\sim$6 AU. The discoveries of these close binary
systems challenge the current planet formation theory with the
question that whether planet formation can go through without
being prevented by the companion's strong perturbations in such close binary
systems.

Current standard theory$-$core accretion model$-$describes the
process of planet formation in 3 phases(Lissauer 1993; Papaloizou
\& Terquem 2006; Armitage 2007): (1) formation  of km-sized
planetesimals($10^{16}$-$10^{20}$ g) from grain coargulation
(Weidenschilling \& Cuzzi 1993) or from gravitational
fragmentation of a dense particle sub-disk near the midplane of
the protoplanetary disk (Goldreich \& Ward 1973), on timescales of
the order of $10^4$ yr; (2) accretion of planetesimals into
planetary embryos ($10^{26}$-$10^{27}$ g, Mercury- to Mars- size)
through a phase of runaway and oligarchic growth on a timescale of
the order of $10^4$-$10^5$ yr (Greenberg et al. 1978; Wetherill \&
Stewart 1989; Barge \& Pellat 1993; Kokubo \& Ida 1996, 1998,
2000; Rafikov 2003, 2004); and (3) giant impacts between embryos,
producing full-sized ($10^{27}$-$10^{28}$ g) terrestrial planets
in about $10^7$-$10^8$ yr (Chambers \& Wetherill 1998; Levison \&
Agnor 2003; Kokubo et al. 2006). For the first
stage (grain-planetesimal phase), a recent study by Pascucci et
al. (2008) suggests that companion stars at projected separation above 10 AU do not appreciably affect the degree of grain
growth. For the later stage(embryo-planet phase), simulations show
binary with periastron of above 10 AU can form terrestrial planets
over the entire range of orbits allowed for single stars(Quintana
\& Lissauer 2007; Quintana et al. 2007). Stellar companions even
hasten the accretion process by stirring up the planetary embryos to
higher eccentricities and inclinations which thus reduce their
orbit-crossing timescales(Quintana 2004; Quintana \& Lissauer
2006). For the intermediate stage(planetesimal-embryo), however,
planetesimal accretion in close binary systems is not as easy as
that around single stars.

The influence of a companion star on the planetesimal accretion
was first studied without including gas drag by Heppenheimer(1974, 1978) and later by Whitmire et al.(1998) . Their results implies that the companion
star, especially if it is on a close orbit with high eccentricity,
may prevent planetesimal accretion by exciting high relative
velocities($\triangle V$) between colliding planetesimals.
$\triangle V$ is crucial parameter, determining whether accretion
or erosion dominates. If $\triangle V$ is greater than the
planetesimal escape velocity [$V_{esc}\sim100R_p/(100$km)
ms$^{-1}$], runaway accretion is prevented and growth is slowed
down. Furthermore, if $\triangle V$ is excited beyond the
threshold velocity $V_{ero}$(a few times of
$V_{esc}$,  depending on the prescription for collision, see the Appendix for details) at which erosion dominates accretion, planetesimal growth ceases.
Fortunately, in the early stages, planetary disk was
rich in gas, which could partly damp companion's perturbations
toward an equilibrium. Furthermore, Marzari \& Scholl (2000) found, in this
equilibrium state, all the orbits of planetesimals with the same size and similar semi-major axes had the configuration of periastron-aligned (so called ``orbital alignment"), hence reducing $\triangle V$ to very low values. Applying this mechanism to the $\alpha$ Cen AB system, accretion zone of km-sized bodies could extend to 2 AU to the primary star ($\alpha$ Cen A). Nevertheless, their
work only took into account the $\triangle V$ of the equal-sized bodies.
Detailed study by Th{\'e}bault et al. (2006) found the orbital
alignment simultaneously increased $\triangle V$ of
different-sized bodies, and it was so efficient
that leaded to significant $\triangle V$ increase for any departure
from the exact equal-sized condition($R_1=R_2$, where $R_1$ and
$R_2$ are the radii of the two colliding bodies). Extending their
studies by including the effect of a fast gas-disk
depletion procedure under UVO photoevaporation from the host star
or nearby stars(Matsuyama et al. 2003; Alexander et al. 2006),
Xie and Zhou (2008) found the effect of differential orbital
alignment could be reduced and all planetesimals
eventually converged toward the same forced orbits, regardless of
size. This mechanism indeed reduces all $\triangle V$ to low
values, however it is not efficient enough for small
bodies(radii less than 5 km) and it take a long time($\sim10^5$-$10^6$ yr) to
reach the ``low $\triangle V"$ state, before which conditions are
still accretion-hostile. Recently, taking into account the differential phasing effect, Th{\'e}bault et al. (2008, 2009) revisited the planetesimal accretion conditions in the $\alpha$ Cen AB system, and found the ``accretion-safe'' area never extended beyond 0.75 AU around Cen A or 0.5 AU aournd Cen B. Their results also confirmed that the planetesimals-to-embryos phase was more affected by the binary environment than the last stages, and it might be one of the biggest difficulties of planet formation in close binary systems.

All previous studies belong to coplanar cases, which have not taken into account
the inclination of the companion's orbit relative to the
planetary disk. According to Hale (1994), coplanarity
is just an approximation with a deviation of the relative orbital inclination of $i_B\le10$ degrees  for solar-type binary systems with separations less than 30-40 AU, and
systems with separations greater than 40 AU appear to have their
orbital planes randomly distributed. Therefore, a small
inclination between the binary orbit and the planetary disk(such
as $i_B$=$1^o$-$5^o$) is more realistic than exact
coplanarity($i_B=0$) for the binary systems with separations of
$\sim 20$ AU.  In this paper, assuming a small inclination for the binary orbit, we reinvestigate the conditions for planetesimal accretion around $\alpha$ Cen A. Results are compared with the previous study Th{\'e}bault et al. (2008). As presented next, by considering the inclination of the companion star, planetesimal accretion is much more favored than in the coplanar case, thus providing a possible solution for the problem of planet formation in close binaries.

In \S 2, we describe our numerical
model and initial conditions. In \S 3, we first compare two nominal cases: $i_B=0$ and $i_B=1$ deg, then present the results of other cases. Some further discussions are  in \S 4. Finally, we summarize in \S 5.

\section{NUMERICAL MODEL AND INITIAL CONDITIONS}
\subsection{Gas Disk Model}
Gas drag is modelled by assuming a non-evolving axisymmetric gas
disk  with constant circular streamlines. Following
Weidenschilling \& Davis (1985), the drag force
 is expressed as
\begin{equation}
   \mathbf{F}={-Kv\mathbf{v}},
\end{equation}
where $\mathbf{F}$ is the force per unit mass, $\mathbf{v}$ is the
relative velocity between the planetesimal and gas, $v$ is the velocity
modulus, and $K$ is the drag parameter, defined as
\begin{equation}
   K={ {3\rho_{g}C_{d}} \over {8\rho_{p} R_{p}} }.
\end{equation}
Here $\rho_{g}$ is the local gas density, and $\rho_{p}$ and $R_{p}$
are the planetesimal density and radius, respectively. $C_{d}$ is a
dimensionless coefficient related to the shape of the body ($\simeq
0.4$ for spherical bodies).

   As we consider the case of $i_B\ne0$, both the distribution and motion of gas should be modeled in the 3-Dimentional space. According to Takeuchi \& Lin (2002), 
in cylindrical coordinates(r, z), we adopt the gas density distribution as
\begin{equation}
   \rho_g(r,z)={ \rho_{g0}({r\over{AU}})^p \exp(-{z^2\over{2h_g^2}})},
\end{equation}
 and the gas rotation as
 \begin{equation}
   \Omega_g(r,z)={ \Omega_{k,mid}[1+{1\over2}({{h_g}\over r})^2(p+q+{q\over2}{z^2\over h_g^2})]},
\end{equation}
where 
$h_g(r)=h_0({r\over{AU}})^{(q+3)/2}$ is the scale height of gas disk, $\Omega_{k,mid}$ is the Keplerian rotation on the mid-plane, and the subscript ``0'' means the value at 1AU. In this paper, we take the Minimum Mass of Solar Nebula(hereafter MMSN, Hayashi 1981) as the nominal gas disk, where $p=-2.75$, $q=-0.5$, $\rho_{g0}=1.4\times10^{-9}$gcm$^{-3}$, and $h_0=4.7\times10^{-2}$ AU.

As stated in many previous studies(Scholl et al. 2007;
Th{\'e}bault et al. 2006), this axisymmetric gas disk model is a
crude simplification, and in reality, perturbations on the gas
disk by the companion should be included(Artymowicz \& Lubow 1994;
Kely \& Nelson 2007). A full description of the dynamical behavior
of planetesimals in such a ``real" disk requires hydro-code
modelling of the gas in addition to N-body type models for
planetesimals. However, there still exist some uncertainties even
in a ``real" disk model: how to choose a numerical wave damping
procedure(Paardekooper et al. 2008).  On the other hand, such an
all-encompassing gas-plus-planetesimals modelling costs much in
calculation and goes beyond the scope of our study in this paper.
Fortunately, the simplified axisymmetric gas disk model, first of
all, is valid on the average; second, it costs very little CPU
time and is thus suited for detailed parameter-space explorations.
Furthermore, as we discuss later, the behavior of $\triangle V$ should probably be similar in both the simplified and full gas models, as long as we have considered the binary inclination. 

\subsection{Initial Set Up}
Table 1 lists most parameters for $\alpha$ Cen AB system, planetesimals and the gas disk. $10^4$ planetesimals with random eccentricities of 0 - 10$^{-4}$, random inclinations of 0 - $5\times10^{-5}$ rad, random semimajor axes of 1 - 2 AU, and random radii of 1 - 10 km, orbit around the primary star, $\alpha$ Cen A. For the nominal gas disk(1MMSN), we explore four cases with different binary inclinations from 0 to 5 deg; for the nominal binary inclination($i_B=1$ deg), we explore three cases with different gas densities(0.1, 1, and 10 MMSN). The inflated radius for collision search is set as $5\times10^{-5}(r_{col}/AU)^{1/2}$ AU, where $r_{col}$ is the distance from a collision to the primary. This choice gives a precision of $\sim$ 1.5 m.s$^{-1}$ in every impact velocity estimate. All the cases evolve $10^4$ yr by using the fourth-order Hermite integrator (Kokubo et al. 1998), including the gas drag force and companion's perturbations.

\section{RESULTS}
\subsection{Coplanar Case vs Inclined Case}
In order to clearly see and fully understand the role of the binary inclination, we present a detail comparison between the nominal coplanar and inclined cases: $i_B=0$ and $i_B =1$ deg. Both cases have a nominal gas disk of 1MMSN.
\subsubsection{Orbital distribution}
In Figure 1, we plot the evolutions of planetesimals' distribution in the X-Z coordinate plane for both $i_B=0$ and $i_B =1$ deg cases. As can be seen in the $i_B=0$ case,  at beginning $t=0$, planetesimals with random radii of  1-10 km are randomly mixed together. As planetesimals move in the gas disk, they are subject to a gas drag force, which progressively damps their orbital inclinations. This inclination damping process is more efficient for small bodies because gas drag force is inversely proportional to planetesimal radius. At the end of the run $t=10^4$ yr, the average Z value of 1 km bodies is almost one order of magnitude lower than those of 10 km ones. Small bodies are crowed in the mid-plane of the disks while big ones are diffused on the surface. For the $i_B=1$ deg case, although with the same initial condition as in the $i_B=0$ case, the evolution of planetesimals' orbits is entirely different. Planetesimals' inclinations are not damped by the gas drag force but excited by the companion star(see the second panels on the right side). 
Under the coupled effect of the companion's secular perturbation and gas drag force, different-sized bodies are eventually forced to orbits with different orbital inclinations.  Because of the size-dependence of gas drag force, smaller planetesimals are damped more by gas drag and thus have smaller forced inclinations than bigger bodies. As can be seen in the two bottom panels on the right side of Figure 1, different-sized bodies are grouped and separated by their radii because of the different forced inclinations. In the next two subsections, we will show that this orbital grouping significantly changes the distribution of impacts among a swam of planetesimals, reduces their impact velocities $\triangle V$ and finally favors planetesimal accretion.

\subsubsection{Impact rate distribution}
As found by Th{\'e}bault et al. (2006), $\triangle V$ is very sensitive to the difference of radii sizes($R_1$ and $R_2$) of the two colliding bodies because of the size-dependence of the orbital alignment. In order to understand the evolution and distribution of $\triangle V$ among a swarm of planetesimals, we, therefore, should first investigate the impact distribution on the $R_1$-$R_2$ plane. In Figure 2, we track the evolution of the impact distributions for both $i_B=0$ and $i_B=1$ deg cases.
In the first $5\times10^2$ yr, the distributions are random and similar in both cases because they start with the same random initial condition and orbital alignments are not fully completed (see Figure 1). Afterwards, the distributions become entirely different: for $i_B=0$ case, impacts occur much more often between bodies of different sizes; for $i_B=1$ deg case, impacts mainly occurs between similar-sized bodies. These two totally different impact distributions can be understood as follows:

i)In the $i_B=0$ case, planetesimal disk is very flattened and can be treat as a 
2-dimensional (2D) disk. Because of orbital alignment in the 2D disk, a planetesimal can collide with another similar-sized one only if their semi-major axes are very close since they have the similar forced orbits (means the similar orbital shapes and configurations) and radial drifts; in contrast, given the different forced orbits and radial drifts, a planetesimal can cross more orbits of other different-sized bodies over a much larger region. 

ii)In the $i_B=1$ deg case, orbital alignment is in 3D space and thus bodies of different sizes are separated from each other (see Figure 1), leading to much lower probability of impact between two planetesimals of different sizes. 

In addition, a big enhancement of impact rate is observed in the $i_B=0$ case since gas drag damps planetesimals' inclinations and thus disk becomes more dense. While in the $i_B=1$ deg case, impact rate is much lower because planetesimals' inclinations are excited and thus the volume density of disk is reduced.

\subsubsection{Accretion vs erosion}
We plot Figure 3 to finally analyze the accretion or erosion conditions for the cases of $i_B=0$ and $i_B=1$ deg. In the two top panels, as expected from the analysis of the  last subsection, the impacts in $i_B=0$ case are much more than those in $i_B=1$ deg case. Since impacts between different-sized bodies are favored in the $i_B=0$ case, the impact velocities are increased to relatively high values with average of $\sim$ 200 ms$^{-1}$; while in the $i_B=1$ deg case, impacts mainly occur between similar-sized bodies with low impact velocities of 10 - 100 ms$^{-1}$ in average. In the two bottom panels, by comparing each $\triangle V$ with two threshold velocities: $V_{low}$ and $V_{high}$ (see the Appendix for detail), we plot the relative weights of the possible types of collision outcomes as a function of distance to the primary star($\alpha$ Cen A). For the case of $i_B=0$, most impacts lead to erosion(red zone). The weight of accretion impact (green + blue zone) is only less than 5\% or about 10\% if the uncertain ones (yellow zone) are included. These results are roughly comparable with the Figure 3 in Th{\'e}bault et al. (2008), which also adopted the assumption of $i_B=0$ but a Maxwellian distribution for planetesimal size. In contrast, for the case of $i_B=1$ deg, accretion is significantly favored: at 1.15 AU,  center of the habitable zone($\sim$1-1.3 AU, e.g. Barbieri et al. 2002), about 40\% (80\%, if including uncertain ones) collisions lead to accretion. If we define the ``possible accretion zone" as green+blue+yellow zones $>$50\%,  it extends to as far as $\sim$2 AU, beyond which companion's perturbations is too strong to reach orbital alignment for planetesimal of any size. Note that our ``possible accretion zone" here is an optimistic definition, where all uncertain ones are taken into account as accretion collision. Actually, the balance between accretion and erosion has a large uncertainty(see the yellow region in Fig.3) because of  the poor constraint on $Q^*$(critical specific energy that leads to catastrophic fragmentation, see the Appendix for detail).

\subsection{Other Cases}
Beside the nominal case with $i_B=1$ deg and gas disk density of 1MMSN, we explore some other cases with different $i_B$ and different gas densities.

\subsubsection{Different $i_B$}
Figure 4 shows the results of two cases with the same gas density of 1 MMSN, but different binary inclinations: $i_B=0.1$ and 5 deg respectively. $i_B=0.1$ deg is marginal case, where every planetesimal has a forced inclination in a very small range 0 $\sim$ 0.1 deg and thus very close to each other. In such case, bodies of different sizes are not well separated from each other and impacts between different-sized bodies are still an important part (see the left up panel of Fig. 4). Therefore, the average impact velocity is intermediate ($\triangle V \sim$ 100 ms$^{-1}$), leading to, although more accretion-friendly than the case of $i_B=0$, still an accretion-hostile condition(red zone exceeds 50\% at least). For the larger inclination case of $i_B=5$ deg, although the planetesimal vertical excursions might actually exceed the typical thickness of the gas disk, the results are similar to the nominal case of $i_B=1$ deg, except for less impact events because of the larger inclination(see the 3 bottom panels in Fig. 4).

\subsubsection{Different gas disks}
We also explore two cases with gas densities of 0.1 MMSN and 10 MMSN in Figure 5. By comparing these two cases with the nominal inclined case in Figure 3, we find: as gas density increases from 0.1 to 10 MMSN, the weight of accretion increases in most outer region but decreases in the inner region(say the habitable zone $\sim$ 1.0-1.3 AU). The probable explanations are: 

i) In the inner region, the companion's perturbations are weak and gas is dense; while in the our region, perturbations are stronger and gas density is lower. As a result of these conditions,  orbital alignment is realized much more easily in the inner than in the outer region.

ii) For 0.1 MMSN gas density, orbital alignment is already realized in the inner region but not in outer region. A larger gas density still favors accretion in the outer region, but it can no longer improve the inner accretion condition. Furthermore, gas drag is a double-edged sword. If orbital alignment is already realized, increasing gas density can result in larger different radial drifts between different-sized bodies, which increase $\triangle V$ and finally prevents accretion. 
\section{DISCUSSION}
\subsection{Behaviors In The Full Gas Model}
As mentioned in \S 2.1, our gas drag model is a simplified one where gas disk is assumed to be fully axisymmetric and non-evolving.  A full gas model requires hydro-code for modeling the gas evolution and N-body type models for planetesimals. Recently, taking the full gas model, Paardekooper et al. (2008) have studied the dynamics and encounter velocities for a swarm of planetesimals in close binaries. 
They found inclusion of the full gas dynamics increased the encounter velocity and made planetesimal accretion even more difficult than in a static, axisymmetric gas disk. Note that their gas disk is two-dimensional, and the binary inclination is also not included. Nevertheless, they have shown some important results that we might infer the $\triangle V$ behavior in the real 3D full gas model, where both the binary inclination and gas evolution are included. As shown in the Figures 9-11 of Paardekooper et al. (2008),  obtial alignments, both of eccentricity and longitude of periastron, are still maintained for bodies of the same size, leading very low $\triangle V$ between equal-sized bodies as in a static, circular gas disk.  This means gas evolution only increase $\triangle V$ between different-sized bodies. In a real 3D full gas model, orbital alignments should be probably maintained, but it is in the 3D space, and thus different-sized bodies are separated from each other, leading to most collisions occurring between similar-sized bodies with low $\triangle V$. Therefore, even in the full gas model, inclusion of a small binary inclination might be also a promising mechanism favoring planetesimal accretion.  The only uncertainty is to what extend can it favor accretion, or whether it is enough to render accretion-friendly impacts dominant? This problem need to be addressed by a future work which can model a 3D evolving gas disk in an inclined binary system.

\subsection{Planetesimal Growth: Type II Runaway Mode}
The major result of this paper is: inclusion of a small binary inclination favors planetesimal accretion significantly, although it might not be enough(if only the green and blue region are counted as accretion) for planetesimal growth to proceed. Even if planetesimal growth can proceed, it cannot be classical single-star-like accretion, or normal accretion. Actually, normal accretion collisions never exceed 20\% in all the cases we investigated. Growth of planetesimals in close binary systems is mainly through ``high-$\triangle V$ accretion" or type II runaway mode (Kortenkamp et al. 2001). 
Type II growth mode is characterised by an initial orderly growth stage followed by a runaway stage. At the beginning, $\triangle V$ is increased and held on values exceeding the biggest objects' escape velocities, but still leading to slow orderly growth. After this orderly phase, growth then switches to runaway when the biggest bodies reach a size large enough for significantly enhancing their gravitational focusing factors: $F_g=(V_{esc}/\triangle V)^2$. This turn-off size directly depends on $\triangle V$. For example, if the runaway growth is turned on at a gravitational focusing factor of $1 - 100$, the biggest objects' escape velocities should be 1- 10 times as large as $\triangle V$. For the coplanar case considered in this paper, the average $\triangle V$ is about 200 m.s$^{-1}$ at 1 AU, thus requiring a turn-off size of 200 - 2000 km. While for the inclined case with $i_B=1$(see Fig. 3), it requires a turn-off size of only 50 -500 km, since the average $\triangle V$ is reduced to 50 m.s$^{-1}$. Therefore, inclusion of the binary inclination, the runaway growth is shifted to an earlier date.

\section{SUMMARY}
In this paper, assuming a small inclination of the binary orbit, we numerical investigate the conditions for planetesimal accretion around $\alpha$ Cen A.
Different-sized (1-10 km randomly) planetesimals are embedded in
the gas disk and moving around the primary from 1 to 2 AU, an extension of its habitable zone($\sim$ 1-1.3 AU). The gas drag is
modeled with a simplified 3-D axisymmetric non-evolving gas disk. Fisrt, we study the effects of the binary inclination on planetesimal accretion by comparing two nominal cases with and without inclination($i_B=0$ and $i_B=1$ deg). Then we explore more cases with different binary inclinations and gas densities. The main results are summarized as follows:

1) In the nominal coplanar case($i_B=0$), planetesimal disk is almost damped to a 2D plane, and thus impact rate is highly increased. While in the nominal inclined case ($i_B=1$ deg), bodies of different sizes are forced to orbits of different inclinations (3D-differential-orbital-phasing) and well separated from each other, leading to much lower total impact rate compared with the coplanar case.

2) In the nominal coplanar case, as differential orbital phasing is almost in a 2D plane, impacts between different-sized bodies are favored, leading to high $\triangle V$ and planetesimal erosion in all the region studied in this paper( 1- 2 AU to the $\alpha$ Cen A). While in the inclined case($i_B=1$ deg), because of the 3D-differential-orbital-phasing, impacts mainly occur between similar-sized bodies, thus leading to much lower $\triangle V$ and a possible accretion zone, which can extend to as far as about 2 AU to the $\alpha$ Cen A when assuming an optimistic $Q^*$ value.

3) In the small inclined case($i_B=0.1$ deg), the results are marginal. Different-sized bodies are not fully separated from each other, and thus a large part of impacts still occur between them, leading to conditions, although much better than that in the coplanar case, still accretion-hostile beyond 1 AU. While for the larger inclined case($i_B=5$ deg), results are comparable with those in the nominal case ($i_B=1$ deg).

4) For the nominal binary inclination($i_B=1$ deg), as gas density increases from 0.1 to 10 MMSN, condition becomes less (more) accretion-hostile in most outer (inner) region.

5) Inclusion of a small binary inclination(say 1-5 deg) significantly favors accretion by reducing $\triangle V$. However, the balance between accretion and erosion still has a large uncertainty, and the reduced $\triangle V$ is not low enough for normal accretion. Planetesimal growth in close binary systems, such as $\alpha$ Cen A,  is probably through ``high-$\triangle V$" collisions or type II runaway mode.

\acknowledgments
We thank Th{\'e}bault, P. for discussions and valuable suggestions.
 This work was supported by  the National Natural Science Foundation of China
  (Nos.10833001 and 10778603) and the National Basic Research Program of China(No.2007CB4800).

\appendix
\section{THRESHOLD VELOCITY}
The threshold velocity of the erosion/accretion limit mainly depends on the critical specific energy $Q^*$. However, different $Q^*$ available in literature often significantly diverge from one another.  As in Figure 8 of Benz \& Asphaug (1999), for example, the difference in $Q^*$ between Benz \& Asphaug (1999) and Holsapple (1994) can be as much as $\sim$2 orders of magnitude in our studied size range(1-10 km). Such an broad $Q^*$ is less usable to interpret our results. Therefore, first, in order to have a clear erosion/limit, we adopt our nominal $Q^*_{nom}=Q^*_{H91}$ as in Housen et al.(1991), second, given the consideration of the uncertainty of $Q^*$, we assume an optimistic $Q^*_{high}=5Q^*_{H91}$. The $Q^*$ in the range of  $Q^*_{H91}-5Q^*_{H91}$ is a reasonable intermediate
value between those in the Holsapple (1994) and Benz \& Asphaug (1999).

Following the detail procedure as in the appendix B of Thebault \& Augereau (2007), we then deduce an ``optimistic threshold velocity" $V_{high}$, by adopting a critical specific energy of $Q^*_{high}$ and a hard excavation coefficient $\alpha=10^{-9}$, and a ``pessimistic threshold velocit'' $V_{low}$, by adopting a critical specific energy of $Q^*_{nom}$ and a weak excavation coefficient $\alpha=4\times10^{-8}$. Figure 6 shows $V_{high}$ and $V_{low}$ vs the radii of target for a projectile of radii of 1 km.

\clearpage
\clearpage
\begin{figure}
\begin{center}
\includegraphics[width=\textwidth]{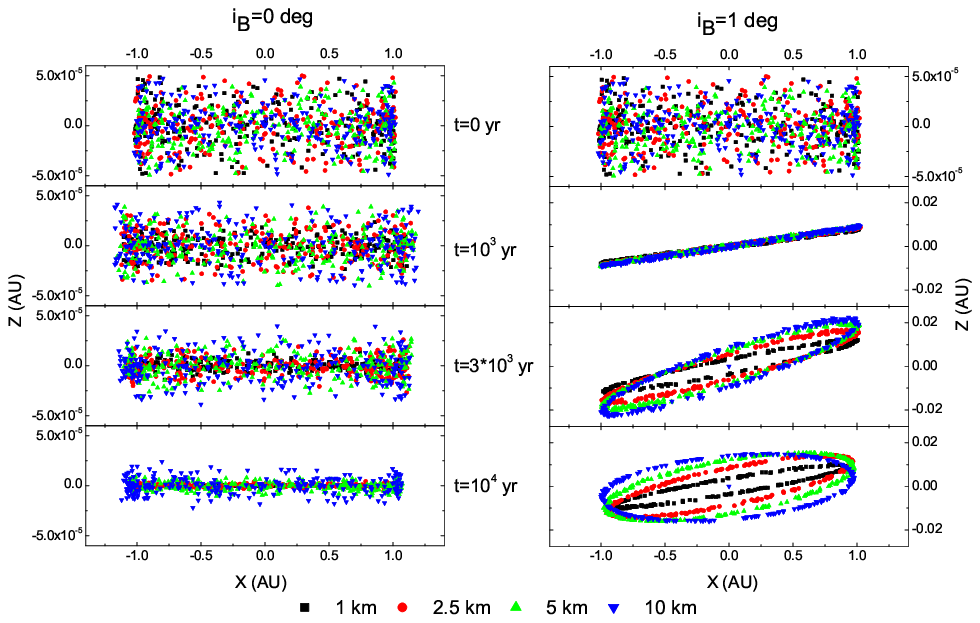}
  \caption{Snapshots of four kinds of size planetesimals(1, 2.5, 5 and 10 km) in the X-Z coordinate plane. Left: nominal coplanar case. Right: nominal inclined case.}
   \end{center}
\end{figure}

\clearpage
\begin{figure}
\begin{center}
\includegraphics[width=\textwidth]{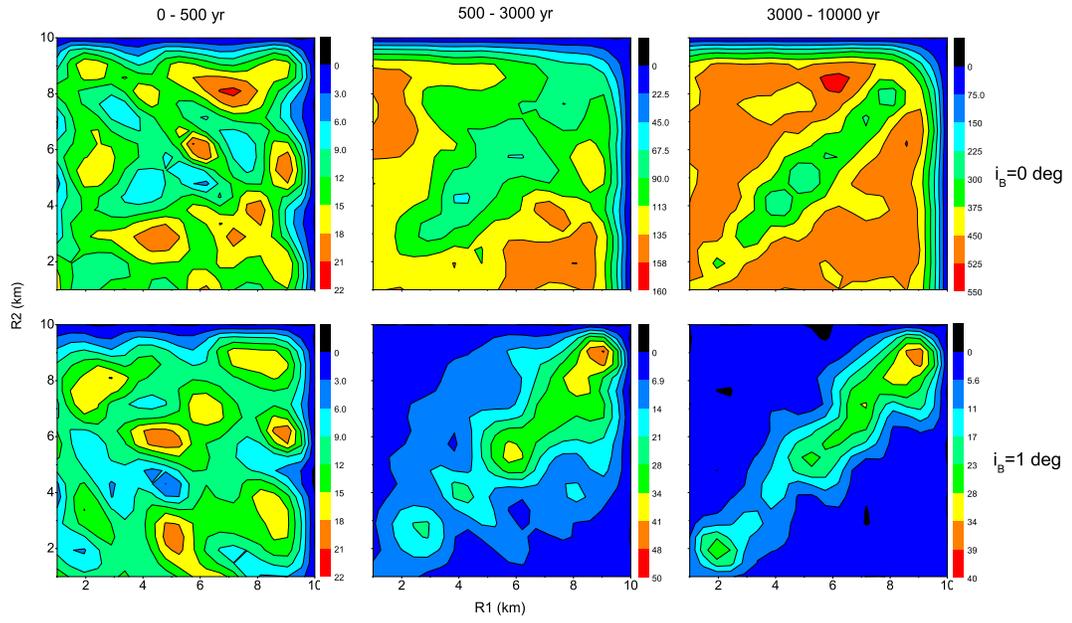}
  \caption{Time-evolution of impact distribution on the $R_1$-$R_2$ plane. Labels of the color bars are the number of impact.}
   \end{center}
\end{figure}

\clearpage
\begin{figure}
\begin{center}
\includegraphics[width=\textwidth]{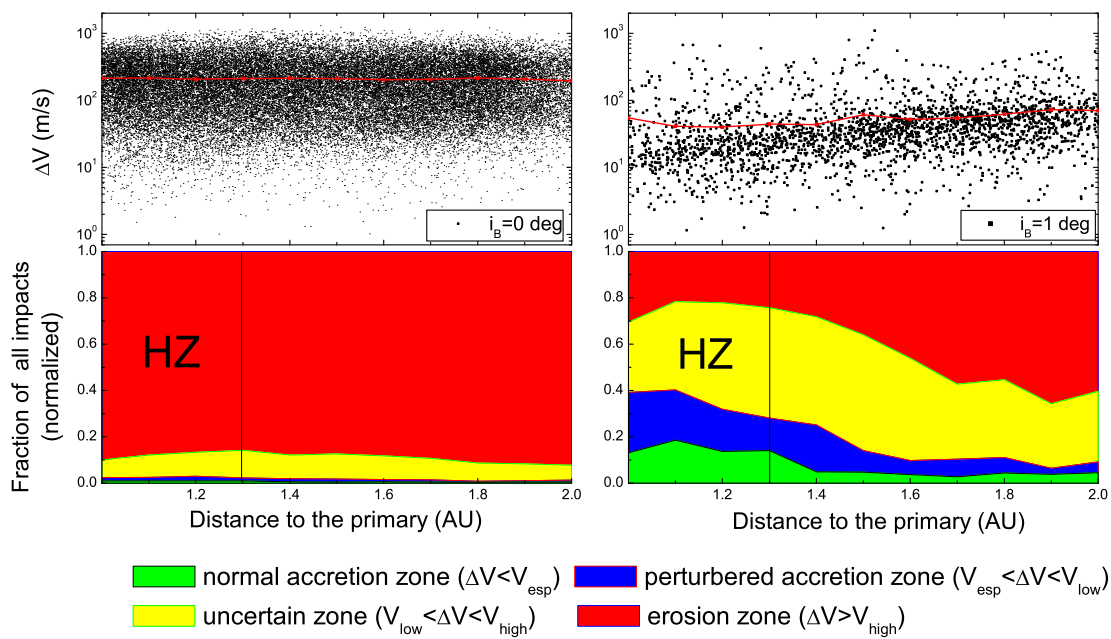}
 \caption{Comparing the accretion conditions for two nominal cases: $i_B=0$ and $i_B=1$ deg. The two top panels: distribution of $\triangle V$, as a function of distance to the primary star. Each black point denotes an collision event, and the red scatter-lines are the average values. The two bottom panels: relative importance of different types of collision outcomes, as a function of distance to the primary star. The vertical lines denote the habitable zone $\sim$ 1-1.3 AU.}
\end{center}
\end{figure}

\clearpage
\begin{figure}
\begin{center}
\includegraphics[width=\textwidth]{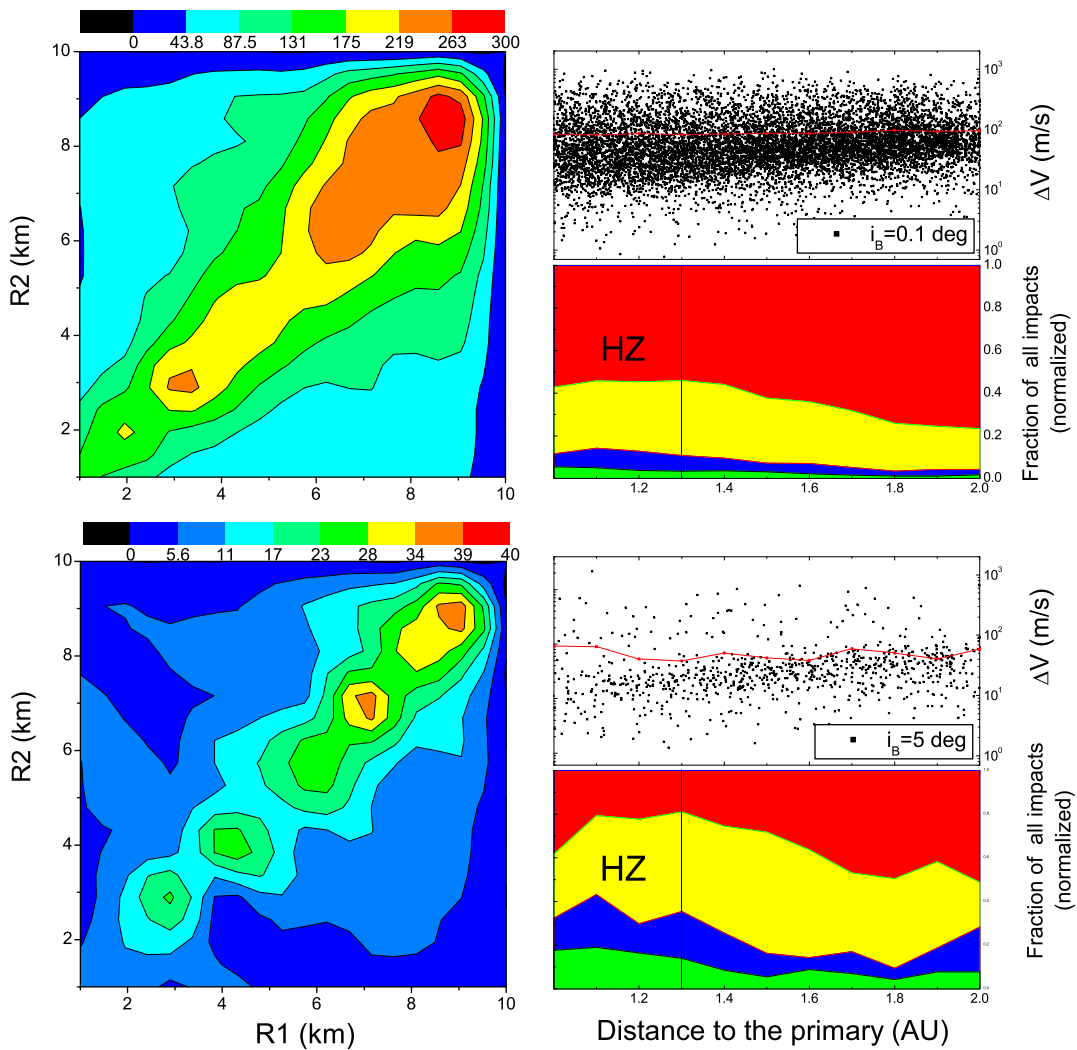}
 \caption{Accretion conditions for two cases with the same gas density of 1 MMSN but different binary inclinations: $i_B=0.1$ deg and $i_B=5$ deg respectively. Left panels: impact distribution on the $R_1$-$R_2$ plane. Right panels: same to figure 3 but for the cases of $i_B=0.1$ deg and $i_B=5$ deg.  }
\end{center}
\end{figure}

\clearpage

\clearpage
\begin{figure}
\begin{center}
\includegraphics[width=\textwidth]{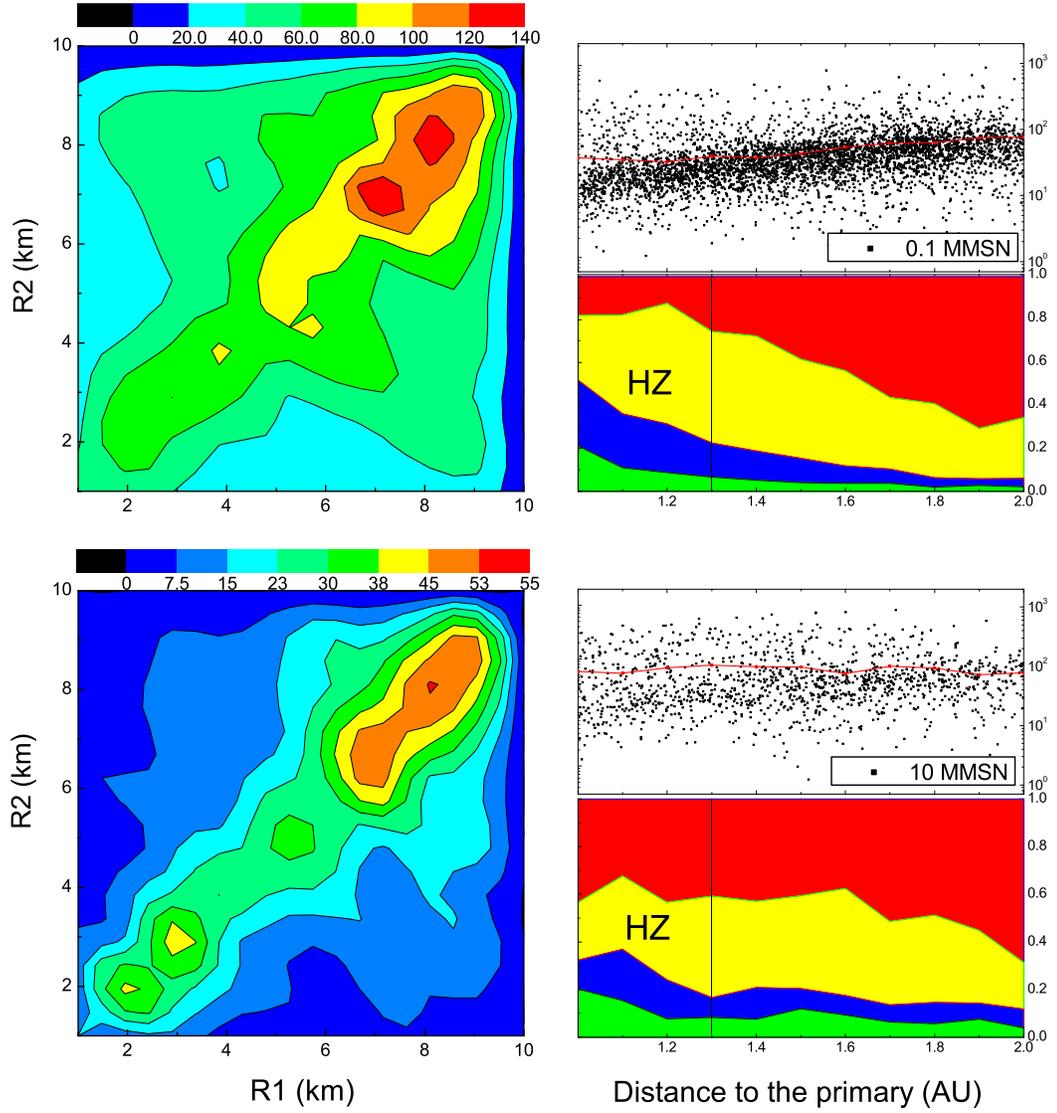}
 \caption{Same to figure 4 but for two cases with different gas densities: 0.1 and 10 MMSN respectively. Both cases have the same binary inclination of 1 deg.}
\end{center}
\end{figure}

\clearpage
\begin{figure}
\begin{center}
\includegraphics[width=\textwidth]{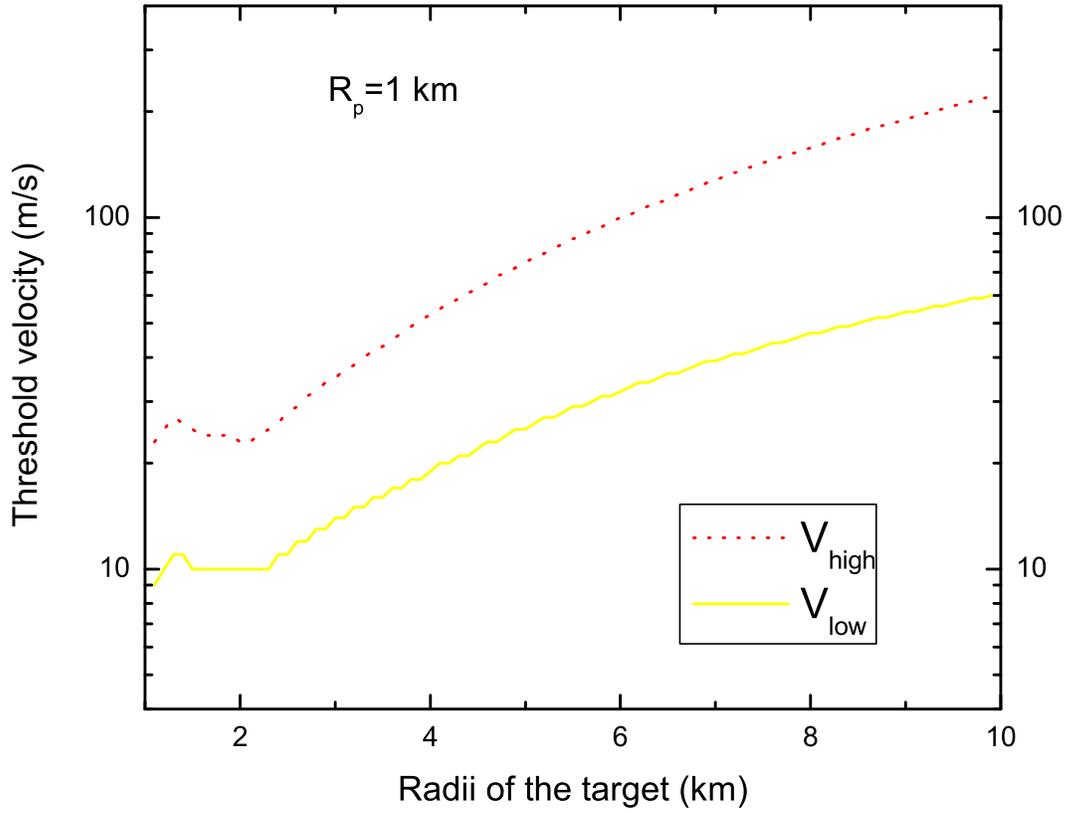}
 \caption{Threshold velocity vs radii of the target with the projectile's radii $R_p=1$ km. Dot: a optimistic estimate of threshold velocity. Solid: a pessimistic estimate of threshold velocity. }
\end{center}
\end{figure}

\clearpage

\clearpage



\begin{table}
 \begin{center}
  \caption{Initial Parameters for the Simulations.\label{tbl-1}}
  \begin{tabular}{@{}llllllllllrrr@{}}
\tableline\tableline
 &Paremeters & &  Values & & Units & &  Descriptions \\
 \tableline
 &$a_{B}$ & &  23.4 & &  $AU$ & &  semi-major axis of companion \\
&$e_{B}$ & &  0.52  & & - & & eccentricity of binary\\
FP*&$i_{B}$ & & 0, 0.1, 1, 5.0  & & $degree$  & & inclination of binary\\
 &$M_{A}$ & &  1.1 & &  $M_{\odot}$ & & mass of primary \\
& $M_{B}$ & &  0.93 & &  $M_{\odot}$ & & mass of companion \\
UD*& $R_{p}$  & & 1 - 10  & & $km$  & & planetesimal radii\\
UD*& $a_{p}$  & & 1.0 - 2.0 & & $AU$  & & semi-major axes of planetesimals \\
UD*&$e_{p}$  & & 0 - $10^{-4}$  & & -  & & eccentricities of planetesimals \\
UD*&$i_{p}$  & & 0 - $5\times10^{-5}$  & & $radian$  & & inclinations of planetesimals \\
& $\rho_{p}$  & & 3.0  & & $g\cdot cm^{-3}$ & & planetesimal density \\
FP*& $\rho_{g0}$ & &  $1.4\times (10^{-10},10^{-9},10^{-8})$  & & $g\cdot cm^{-3}$  & & initial gas density at 1AU \\
& $\rho_{g}$ & &  $r^{-11/4}$ & &  - & &  initial gas density radial profile \\
& $N_p$ & &  $10^4$  & &  -  & & number of test particles\\
& $R_{inflated}$ & &  $5\times10^{-5}(r_{col}/AU)^{1/2}$  & &  $AU$  & & Inflated radius for collision search\\
& $t_{final}$ & &  $10^4$ & &  $yr$  & & final timescale for each run\\

\tableline
\end{tabular}
\tablecomments{$M_{\odot}$ stands for the solar mass. The fields marked by FP* are explored as free parameters in the simulations, while the ones marked by UD* have an uniform distribution.}
\end{center}
\end{table}




\begin{thebibliography}{99}
\bibitem[\protect\citeauthoryear{Adachi}{1976}]{b1}
Adachi, I., Hayashi, C., \& Nakazawa, K. 1976, Prog. Theor. Phys.,
56, 1756

\bibitem[\protect\citeauthoryear{Alexander}{2006}]{b2}
Alexander, R. D., Clarke, C. J., Pringle, J. E. 2006b, MNRAS, 369,
229

\bibitem[\protect\citeauthoryear{Armitage}{2007}]{b3}
Armitage, P. J. 2007, Lecture note on the formation and evolution of
planetary systems, Preprint (astro-ph/0701485)

\bibitem[\protect\citeauthoryear{Artymowicz}{1994}]{b4}
Artymowicz, P. \& Lubow, S.H., 1994, AJ 421, 651

\bibitem[\protect\citeauthoryear{Barbieri}{2002}]{b5}
Barbieri, M., Marzari, F., Scholl, H., 2002, A\&A, 396, 219

\bibitem[\protect\citeauthoryear{Barge}{1993}]{b6}
Barge, P., \& Pellat, R. 1993, Icarus, 104, 79

\bibitem[\protect\citeauthoryear{Benz}{1999}]{b7}
Benz, W., \& Asphaug, E., 1999, Icarus, 142, 5

\bibitem[\protect\citeauthoryear{Chambers}{1998}]{b8}
Chambers, J. E., \& Wetherill, G. W. 1998, Icarus, 136, 304

\bibitem[\protect\citeauthoryear{Duch{\^e}ne}{2007}]{b8}
Duch{\^e}ne, G., DelgadoÐDonate, E., Haisch,
K. E., Jr., Loinard, L., Rodr«õguez, L. F.
2007, 379, in Protostars and Planets V,
eds. B. Reipurth, D. Jewitt, \& K. Keil
(Tucson: Univ. Arizona Press)

\bibitem[\protect\citeauthoryear{Duquennoy}{1991}]{b9}
Duquennoy, A. \& Mayor, M. 1991, A\&A,
248, 485

\bibitem[\protect\citeauthoryear{Eggenberger}{2004}]{b10}
Eggenberger, A., Udry, S., \& Mayor, M. 2004, A\&A, 417, 353

\bibitem[\protect\citeauthoryear{Eggenberger}{2007}]{b11}
Eggenberger, A., Udry, S., Mazeh, T., Segal, Y., \& Mayor, M. 2007, A\&A, 466, 1179

\bibitem[\protect\citeauthoryear{Goldreich}{1973}]{b12}
Goldreich, P., \& Ward, W. R. 1973, ApJ, 183, 1051

\bibitem[\protect\citeauthoryear{Greenberg}{1978}]{b13}
Greenberg, R., Hartmann, W. K., Chapman, C. R., Wacker, J. F. 1978,
Icarus, 35, 1

\bibitem[\protect\citeauthoryear{Greenzweig}{1992}]{b14}
Greenzweig, Y., \& Lissauer, J. J., 1992, Icarus, 100, 440

\bibitem[\protect\citeauthoryear{Hale}{1994}]{b15}
Hale, A. 1994, AJ, 107, 306

\bibitem[\protect\citeauthoryear{Hatzes}{2003}]{b16}
Hatzes, A.P., Cochran, W.D., Endl, M., McArthur, B., Paulson, D.,
Walker, G.A.H., Campbell, B., \& Yang, S. 2003, ApJ, 599, 1383

\bibitem[\protect\citeauthoryear{Heppenheimer}{1974}]{b17}
Heppenheimer, T. A. 1974, Icarus, 22, 436

\bibitem[\protect\citeauthoryear{Heppenheimer}{1978}]{b18}
$-$ 1978, A\&A, 65, 421

\bibitem[\protect\citeauthoryear{Holsapple}{1994}]{b19}
Holsapple, K., 1994, Planet. Space Sci., 42, 1067

\bibitem[\protect\citeauthoryear{Housen}{1991}]{b20}
Housen, K., Schmidt, R. M., \& Holsapple, K., 1991, Icarus, 94, 180

\bibitem[\protect\citeauthoryear{Kley}{2007}]{b21}
Kley, W., \& Nelson, R. 2007, On the Formation and Dynamical
Evolution of Planets in Binaries, preprint (astro-ph/07053421)

\bibitem[\protect\citeauthoryear{Kokubo}{1996}]{b22}
Kokubo, E. \& Ida, S. 1996, Icarus, 123, 180

\bibitem[\protect\citeauthoryear{Kokubo}{1998}]{b23}
$-$ 1998, Icarus, 131, 171

\bibitem[\protect\citeauthoryear{Kokubo}{2000}]{b24}
$-$ 2000, Icarus, 143, 15

\bibitem[\protect\citeauthoryear{Kokubo}{2006}]{b25}
Kokubo, E., Kominami, J., Ida, S. 2006, ApJ, 642, 1131

\bibitem[\protect\citeauthoryear{Kokubo et al.}{1998}]{b26}
Kokubo, E., Yoshinaga K., \& Makino, J. 1998, MNRAS, 297, 1067

\bibitem[\protect\citeauthoryear{Konacki}{2005}]{b27}
Konacki, M. 2005, Nature, 436, 230

\bibitem[\protect\citeauthoryear{Kortenkamp}{2001}]{b28}
Kortenkamp, S.~J., Wetherill, G.~W., \& Inaba, S.  2001, Science, 293, 1127

\bibitem[\protect\citeauthoryear{Levison}{2003}]{b29}
Levison, H. F., \& Agnor, C. 2003, AJ, 125, 2692

\bibitem[\protect\citeauthoryear{Lissauer}{1993}]{b30}
Lissauer, J. J. 1993, ARA\&A, 31, 129

\bibitem[\protect\citeauthoryear{Marzari}{2000}]{b31}
Marzari, F. \& Scholl, H. 2000, ApJ, 543, 328

\bibitem[\protect\citeauthoryear{Mathieu}{2000}]{b32}
Mathieu, R. D., Ghez, A. M., Jensen, E.
L. N., Simon, M. 2000, 703, in Protostars
and Planets IV, eds. Mannings, V.,
Boss, A.P., Russell, S. S. (Tucson: Univ. Arizona Press)

\bibitem[\protect\citeauthoryear{Matsuyama}{2003}]{b33}
Matsuyama, I., Johnstone, D., \& Hartmann, L. 2003, ApJ, 582, 893

\bibitem[\protect\citeauthoryear{Mugrauer}{2005}]{b34}
Mugrauer, M., Neuh¬auser, R., Seifahrt, A., Mazeh, T., \& Guenther, E. 2005,
A\&A, 440, 1051

\bibitem[\protect\citeauthoryear{Paardekooper}{2008}]{b35}
Paardekooper, S.-J., Th{\'e}bault, P., \& Mellema, G.\ 2008, \mnras, 386, 973

\bibitem[\protect\citeauthoryear{Papaloizou}{2006}]{b36}
Papaloizou, J. C. B., Terquem, C. 2006, Rep. Prog. Phys., 69, 119

\bibitem[\protect\citeauthoryear{Pascucci}{2008}]{b37}
Pascucci, I., Apai, D., Hardegree-Ullman, E.~E., Kim, J.~S., Meyer, M.~R.,
\& Bouwman, J.\ 2008, \apj, 673, 477

\bibitem[\protect\citeauthoryear{Queloz}{2000}]{b38}
Queloz, D., et al. 2000, A\&A, 354, 99

\bibitem[\protect\citeauthoryear{Quintana}{2004}]{b39}
Quintana, E.~V., 2004, Ph.D. thesis, Univ. Michigan

\bibitem[\protect\citeauthoryear{Quintana}{2006}]{b40}
Quintana, E. V., \& Lissauer, J. J. 2006, Icarus, 185, 1

\bibitem[\protect\citeauthoryear{Quintana}{2007}]{b41}
Quintana, E.~V., Adams, F.~C., Lissauer, J.~J., \& Chambers, J.~E.\ 2007, \apj, 660, 807

\bibitem[\protect\citeauthoryear{Quintana}{2007}]{b42}
Quintana, E. V. \& Lissauer, J. J. 2007,
in Planets in Binary Star Systems, ed.
N. Haghighipour (Springer Publishing Company)

\bibitem[\protect\citeauthoryear{Rafikov}{2003}]{b43}
Rafikov, R. R. 2003, AJ, 125, 942

\bibitem[\protect\citeauthoryear{Rafikov}{2004}]{b44}
Rafikov, R. R. 2004, AJ, 128, 1348

\bibitem[\protect\citeauthoryear{Scholl}{2007}]{b45}
Scholl, H., Marzari, F., Th{\'e}bault, P. 2007, MNRAS, 380, 1119

\bibitem[\protect\citeauthoryear{Takeuchi}{2002}]{b46}
Takeuchi, T. \& Lin, D. N. C., 2002, \apj, 581, 1344

\bibitem[\protect\citeauthoryear{Th{\'e}bault}{2007}]{b47}
Th{\'e}bault, P., \& Augereau, J.-C., 2007, A\&A, 472, 169

\bibitem[\protect\citeauthoryear{Th{\'e}bault}{2006}]{b48}
Th{\'e}bault, P., Marzari, F., \& Scholl, H. 2006, Icarus, 183, 193

\bibitem[\protect\citeauthoryear{Th{\'e}bault}{2008}]{b49}
$-$, 2008, \mnras, 388, 1528

\bibitem[\protect\citeauthoryear{Th{\'e}bault}{2009}]{b50}
$-$, 2009, \mnras, 393, L21

\bibitem[\protect\citeauthoryear{Weidenschilling}{1993}]{b51}
Weidenschilling, S. J., \& Cuzzi, J. N. 1993, in Protostars and
Planets III, ed. E. H. Levy \& J. I. Lunine (Tucson: Univ. Arizona
Press), 1031

\bibitem[\protect\citeauthoryear{Weidenschilling}{1985}]{b52}
Weidenschilling, S. J. \& Davis, D. R. 1985, Icarus, 62, 16

\bibitem[\protect\citeauthoryear{Wetherill}{1989}]{b53}
Wetherill, G. W., Stewart, G. R. 1989, Icarus, 77, 330

\bibitem[\protect\citeauthoryear{Wetherill}{1993}]{b54}
$-$. 1993, Icarus, 106, 190

\bibitem[\protect\citeauthoryear{Whitmire}{1998}]{b55}
Whitmire, D. P., Matese, J. J. \& Criswell L. 1998, Icarus, 132, 196

\bibitem[\protect\citeauthoryear{Xie}{2008}]{b56}
Xie, Ji-Wei, \& Zhou, Ji-Lin,  2008, \apj, 686, 570\

\bibitem[\protect\citeauthoryear{Zucker}{2004}]{b57}
Zucker, S., Mazeh, T., Santos, N. C., Udry, S., \& Mayor, M. 2004, A\&A, 426,
695
\end{thebibliography}
\end{document}